\RequirePackage{fix-cm}
\documentclass[twocolumn]{svjour3}          
\smartqed  
\usepackage{graphicx}
\usepackage[backend=biber, style=phys, sorting=none, doi=true, eprint=true, biblabel=brackets,maxbibnames=4]{biblatex}
\usepackage{csquotes}

\AtEveryBibitem{%
  \clearfield{Extra}%
}

\AtEveryBibitem{%
  \clearfield{note}%
}

\AtEveryBibitem{%
  \clearfield{series}%
}

\AtEveryBibitem{%
  \clearfield{doi}%
}

\AtEveryBibitem{%
  \clearlist{language}%
}

\usepackage{textalpha}
\DeclareUnicodeCharacter{2212}{-}
\DeclareUnicodeCharacter{2009}{ }
\DeclareUnicodeCharacter{2248}{$\approx$}
\DeclareUnicodeCharacter{202F}{ }
\DeclareUnicodeCharacter{223C}{ }
\DeclareUnicodeCharacter{2086}{$_6$}
\DeclareUnicodeCharacter{2084}{$_4$}
\DeclareUnicodeCharacter{2082}{$_2$}
\DeclareUnicodeCharacter{2083}{$_3$}

\usepackage{amsmath,amssymb}
\usepackage{cuted}
\usepackage{flushend}
\AfterEndEnvironment{strip}{\leavevmode}
\usepackage{silence}
	\ErrorsOff[latex]

\DefineBibliographyStrings{english}{
	andothers = {\mkbibemph{et al.}}
}

\usepackage{xcolor,color,soul}

\AtBeginBibliography{\small}

\begin{document}

\title{On the lineshapes of temperature-dependent transport measurements of superconductors under pressure
}

\author{Alexander C. Mark         \and
        Russell J. Hemley 
}

\institute{Alexander C. Mark \at
              Department of Physics \\
              University of Illinois Chicago\\
              Chicago, Illinois 60607\\
           \and
           Russell J. Hemley \at
              Departments of Physics, Chemistry, and Earth and Environmental Sciences \\
              University of Illinois Chicago\\
              Chicago, Illinois 60607\\
}

\date{Received: date / Accepted: date}


\begin{strip}
\begin{center}
{\Large \textbf{On the lineshapes of temperature-dependent transport \\ measurements of superconductors under pressure} \vspace{10pt}}
\\

{Alexander C. Mark$\mathrm{^a}$ and Russell J. Hemley$\mathrm{^b}$ \vspace{5pt}} 
\\

$\mathrm{^a}$ \textit{Department of Physics, University of Illinois Chicago, Chicago IL 60607}\\
$\mathrm{^b}$ \textit{Departments of Physics, Chemistry, and Earth and Environmental Sciences, \\ University of Illinois Chicago, Chicago IL 60607}
\end{center}
\end{strip}

\begin{abstract}
Recent reports of superconductivity in the vicinity of room temperature have been the subject of discussion by the community. Specifically, features in the resistance-temperature ($R$-$T$) relations have raised questions. We show that many of these features can arise from previously unaccounted-for dynamic effects associated with the AC transport techniques often used in high-pressure experiments. These dynamic AC effects can can cause the apparent resistance ($\mathit{R_{apparent}}$) to diverge from the DC resistance ($\mathit{R_{DC}}$), sharpen measured superconducting transitions, and produce other features in the measured $R$-$T$ response. We also show that utilizing the full output of phase-sensitive transport measurements provides a valuable probe of superconducting samples in difficult to measure systems.

\end{abstract}

\section{Introduction}
\label{intro}
The discovery of very high-temperature superconductivity in pressurized hydrides has been a major milestone in the route to ambient (\textit{i.e.}, room temperature and pressure) superconductivity.  Discoveries include superconducting critical temperatures ($T_c$) above 200 K in $\mathrm{H_3S}$ \cite{drozdov_conventional_2015}, $\mathrm{LaH_{10}}$ \cite{somayazulu_evidence_2019,drozdov_superconductivity_2019,hong_superconductivity_2020}, $\mathrm{YH_9}$ \cite{kong_superconductivity_2021,snider_synthesis_2021}, $\mathrm{YH_6}$ \cite{troyan_anomalous_2021}, and $\mathrm{CaH_6}$ \cite{ma_high-temperature_2022,li_superconductivity_2022} and a growing number of ternary hydrides such as C-S-H \cite{pasan_observation_nodate}, La-Y-H \cite{semenok_superconductivity_2021} and La-Al-H \cite{chen_high-temperature_2024}, many of which were theoretically predicted \cite{wang_superconductive_2012, li_metallization_2014, duan_pressure-induced_2014, liu_potential_2017, peng_hydrogen_2017}. Understanding the origin and nature of superconductivity in these materials requires detailed analyses of electrical resistance, magnetic susceptibility, and other properties of these systems. Of particular interest have been features present in the $R$-$T$ data near, above, and below $T_c$ reported in different studies \cite{drozdov_conventional_2015,somayazulu_evidence_2019, drozdov_superconductivity_2019, hong_superconductivity_2020, kong_superconductivity_2021, snider_synthesis_2021,  ma_high-temperature_2022, li_superconductivity_2022, pasan_observation_nodate, semenok_superconductivity_2021, chen_high-temperature_2024, mozaffari_superconducting_2019, dias_observation_2023, salke_evidence_nodate}.

Superconductors are characterized using both AC or DC transport methods \cite{tinkham_introduction_2004}. AC techniques, including phase-sensitive detection methods, are used to enhance the signal-to-noise ratio (SNR) when measurement signals are very low and often buried within noise or background signals. As a result, AC methods are often used in high-pressure experiments, especially at very high (\textit{e.g.}, megabar) pressures in diamond-anvil cells where sample sizes and therefore signal levels are necessarily small. While the superconducting $R$ drop at $T_c$ may be intrinsically sharp (\textit{e.g.}, 2D superconductors \cite{guo_crossover_2020}), the measured sharpness of the transition, as well as the temperature dependence in the normal state, may also be influenced by the parameters of the measurement,as pointed out in the original measurements for $\mathrm{LaH_{10}}$ \cite{somayazulu_evidence_2019}. Being sensitive to both resistive and reactive effects, AC techniques can introduce frequency dependent features in the measured response that are not observed with DC probes \cite{stuller_introduction_2007}.

\begin{figure*}
  \includegraphics[width=\textwidth]{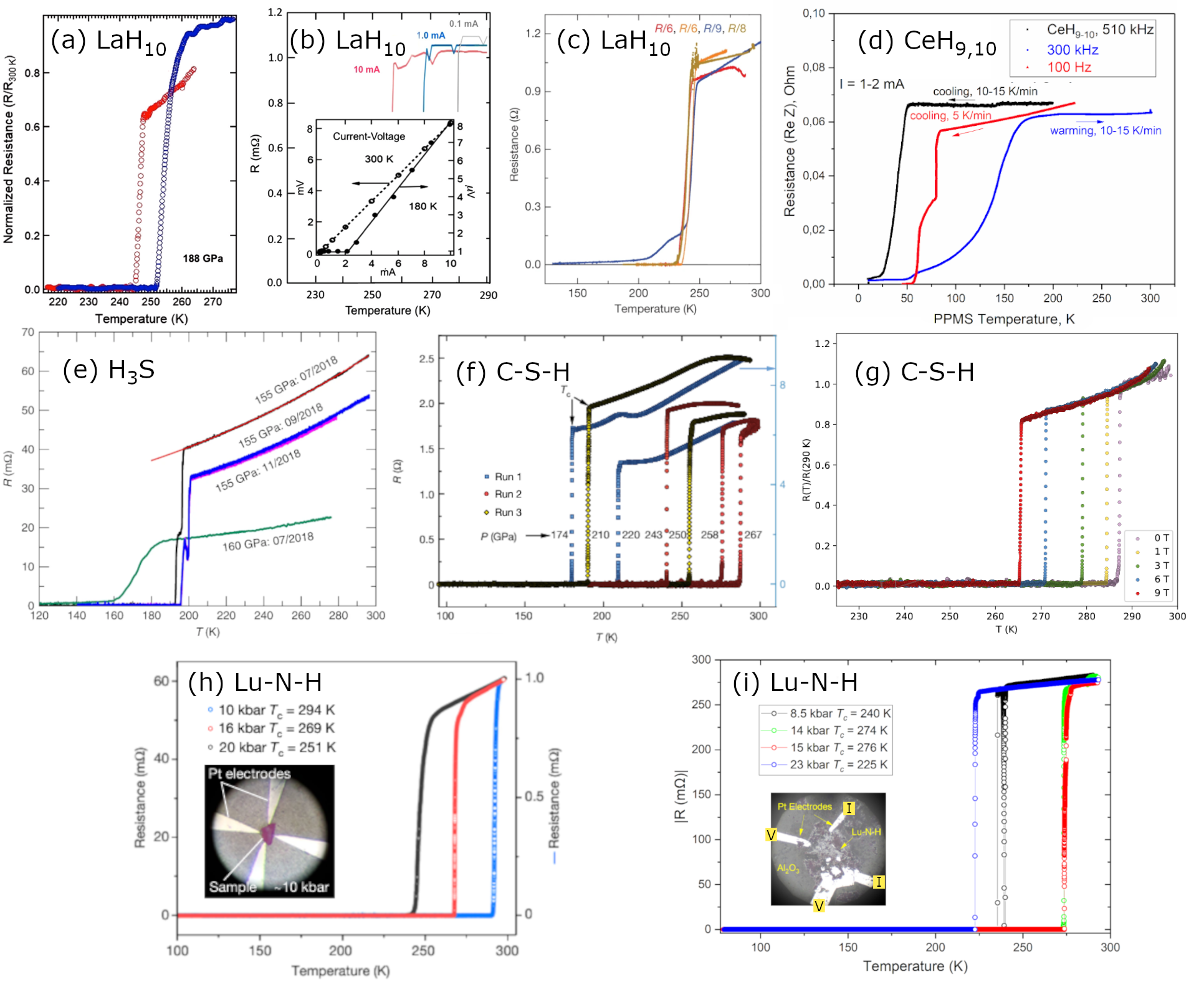}
\caption{Examples of $\mathit{R}$-$\mathit{T}$ relations of hydride superconductors exhibiting features above and below $\mathit{T_c}$ that have been discussed in the literature. a) and b) $\mathrm{LaH_{10}}$ \cite{somayazulu_evidence_2019}, c) $\mathrm{LaH_{10}}$ \cite{drozdov_superconductivity_2019}, d) $\mathrm{CeH_{9,10}}$ \cite{semenok_evidence_nodate}, e) $\mathrm{H_3S}$  \cite{mozaffari_superconducting_2019}, f) C-S-H (from Ref. \cite{hirsch_enormous_2023}), g) C-S-H (from Ref. \cite{hamlin_vector_nodate}), h) Lu-N-H \cite{dias_observation_2023}, i) Lu-N-H  \cite{salke_evidence_nodate}. Figures adapted from the references above.}
\label{Fig:collage}       
\end{figure*}

Figure \ref{Fig:collage} shows examples of apparent narrowing of the superconducting transition, peaks above $\mathit{T_c}$, two-step resistance drops, and non-zero offsets below $T_c$ in samples measured with AC transport techniques \cite{somayazulu_evidence_2019, drozdov_superconductivity_2019, mozaffari_superconducting_2019, hirsch_enormous_2023, hamlin_vector_nodate, dias_observation_2023, salke_evidence_nodate,semenok_evidence_nodate} . We show that these features, many of which have been subjects of discussion in the literature  \cite{hirsch_nonstandard_2021,dogan_anomalous_2021,dias_standard_nodate,hamlin_vector_nodate, harshman_analysis_nodate,talantsev_broadening_nodate,hirsch_enormous_2023}, can arise from the specific characteristics of the AC detection methods used in these high-pressure experiments. These features can be understood by considering simple electrodynamics, electrical circuit concepts, and signal processing effects that are not typically presented in published reports. 

\section{Modeling AC Transport Measurements}
\label{modeling measurements}

A common AC measurement technique relies on phase-sensitive detection, \textit{i.e.}, lock-in amplification. Lock-in measurements can be modeled using standard in-phase ($\mathit{X}$), quadrature ($\mathit{Y}$), amplitude ($|\mathit{V}|$), and phase angle ($\theta$) output signals from the amplifier. Mathematical definitions for each are provided in the Supplemental Information. The discussion will concern parasitic resistances, capacitances, and inductances that contribute to the signal. These parasitic effects are expected to be most prominent in small and/or inhomogeneous samples \cite{beltran_effect_2003,farmehini_-chip_2020}.

\subsection{\textbf{Specific Experimental Setups}}
\label{sub:setups}
\begin{figure}
\centering
  \includegraphics[width=0.5\textwidth]{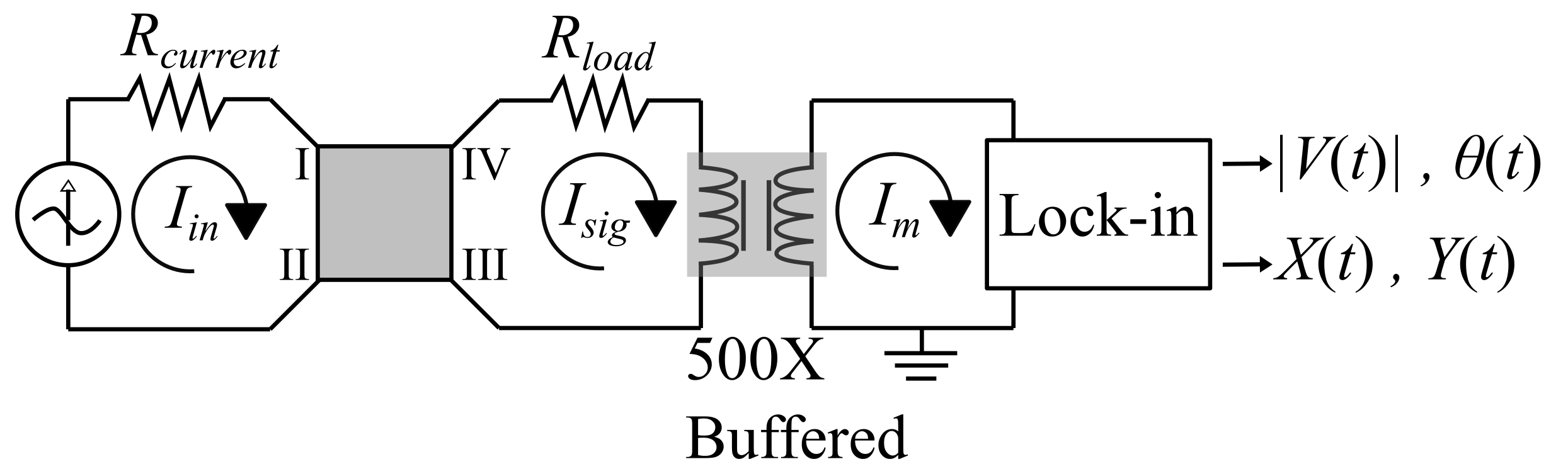}
\caption{Example of the schematic block diagram of a setup used for small-signal transport measurements. Small parasitic resistances in the current and voltage lines are represented by $\mathit{R_{current}}$ and $\mathit{R_{load}}$, respectively. All equipment in physical contact with the sample is electrically floating to ensure large values of $\mathit{R_{load}}$ (\textit{e.g.} a failing contact) do not cause the pre-amplifier to output zero volts due to the sample being shunted to ground (see Ref. \cite{noauthor_model_2004} and Supplemental Information).}
\label{Fig:Probes}       
\end{figure}

Transport measurements at high pressures are typically conducted using a four-probe geometry, sometimes referred to as a van der Pauw (VDP) configuration. It is important to note that while a VDP geometry is often stated for these experiments, this does not imply the VDP algorithm \cite{van_der_pauw_method_1958} was used to determine the sheet resistivity of a sample unless specifically stated. The majority of the measurements exhibiting unusual features were not performed using the VDP method, but instead utilized techniques based on AC bridge circuitry \cite{snider_synthesis_2021,cross_high-temperature_2024} or a boxcar-averaged DC circuit \cite{drozdov_conventional_2015, drozdov_superconductivity_2019} (\textit{e.g.}, Quantum Design PPMS options \cite{noauthor_physical_nodate}). Other studies reported experiments using lock-in-based techniques with circuitry similar to that shown in Fig. \ref{Fig:Probes} \cite{somayazulu_evidence_2019, grockowiak_hot_2022,cross_high-temperature_2024}. As all of the aforementioned techniques rely on periodic averaging, they are inherently susceptible to dynamic effects arising from sample inductance and capacitance. The relatively small size and granular nature of hydride superconductors often cause the typically negligible frequency-dependent contribution to the measured signals to approach the magnitude of the resistive contribution \cite{troyan_high-temperature_2022, eremets_high-temperature_2022}. 

In a lock-in measurement, current-regulated sinusoidal signals of the form 
\begin{equation}
\mathit{I_{in}(t)}=|\mathit{I_{max}}|\sin (\mathit{\omega t})
\end{equation} 
is injected into the sample through two points (I and II) by a regulated current source (Fig. \ref{Fig:Probes}). A differential voltage of the form 
\begin{equation}
\mathit{V_{out}(t)}=|\mathit{V}| \sin (\mathit{\omega t} + \theta)
\end{equation} 
is then measured across the other two contacts (III and IV), buffered via an active pre-amplifier, and then measured by a lock-in amplifier. Before taking any measurements, it must be ensured that excitation frequencies are within the pass band of all circuitry in the setup. 

Typically, effective in-phase and quadrature resistances are then defined in terms of $\mathit{X(t)}$ and $\mathit{Y(t)}$ using Ohms law as 
\begin{equation}
\mathit{R_{ip}(t,\omega)} = \frac{\mathit{X(t,\omega)}}{\mathit{I(t,\omega)}}
\end{equation} 
and 
\begin{equation}
\mathit{R_{qd}(t,\omega)} = \frac{\mathit{Y(t,\omega)}}{\mathit{I(t,\omega)}},
\end{equation}
respectively. A complex impedance, 
\begin{equation}
\mathit{\mathbf{Z}(t,\omega)} = \mathit{R_{ip}(t,\omega)}+\mathit{iR_{qd}(t,\omega)}=|\mathit{Z(t,\omega)}| \ e^{\mathit{i}\theta(\mathit{t,\omega)}}
\end{equation} 
is then used to estimate $\mathit{R_{DC}}$. For ohmic materials $\mathit{R_{ip}(t,\omega)} >> \mathit{R_{qd}(t,\omega)}$, so the apparent resistance is estimated as $\mathit{R_{apparent}(t,\omega)} = |\mathit{Z(t,\omega)}| \approx \mathit{R_{DC}(t)}$ for a wide range of conditions. While the voltage across ideal ohmic materials is expected to remain in phase with the excitation current, all real materials exhibit dynamic electromagnetic effects with at least a diamagnetic effect present \cite{jackson_john_2015}. For non-ohmic materials, such as superconductors, the actual $I$-$V$ relation may have a significant $\mathit{\omega}$ dependence and this approximation is invalid \cite{kaiser_electromagnetic_2004, brown_engineering_2006}. 

To understand how these dynamic effects contribute to the measured $\mathit{X}$ and $\mathit{Y}$, the time-dependent $I$-$V$ relation may be written 
\begin{equation}
|\mathit{V(t,\omega)}| \ \mathrm{sin}(\mathit{\omega t} + \theta) = \mathit{Z(t,\omega)} \ |\mathit{I}| \ \mathrm{sin}(\mathit{\omega t})
\end{equation}
\noindent where $\mathit{Z(\omega,t)}$ is the frequency-dependent impedance. This relation is then converted to phasor notation:
\begin{equation}
\mathit{\mathbf{V}(t,\omega)}=\mathit{\mathbf{I}(t,\omega)}\mathcal{H}\mathit{(t,\omega)},
\end{equation} 
where $\mathit{\mathbf{V}(t,\omega)}$ is a complex sinusoidal voltage signal and $\mathit{\mathcal{H}(t,\omega)}$ is a complex transfer function that is determined by the physical properties of the sample \cite{stuller_introduction_2007,gates_introduction_2001}.

The presence of a finite offset from zero resistance in the $\mathit{R_{apparent}}$ of samples reported to be superconducting measured with AC techniques has been the subject of frequent discussion (\textit{e.g.}, Refs. \cite{hamlin_vector_nodate,harshman_analysis_nodate}). Here it is important to note that the use of $|\mathit{V(t)}|$, being a vector magnitude, to estimate $\mathit{R_{DC}}$ will result in a scalar voltage offset. While lock-in amplifiers can extract extremely small signals from background noise, there is still a finite noise floor in any real measurement. For extremely low voltage signals, \textit{e.g.}, zero voltage drop in a superconductor below $\mathit{T_c}$, the signal is guaranteed to be below the lock-in noise floor of the amplifiers. It is naively expected that using Ohm's law to calculate the temperature dependence of the DC resistance ($\mathit{R_{DC}(T)}$) based on $|\mathit{V(T,\omega)}|$ will result in $\mathit{R_{apparent}(T<T_c)} \approx 0 $ with environmental and instrumental noise causing individual readings to symmetrically fluctuate around $0$. This discrepancy is addressed by recognizing that the lock-in amplifier only measures $\mathit{X(T)}$ and $\mathit{Y(T)}$, with both $|\mathit{V(T)}|$ and $\theta(\mathit{T})$ being calculated in hardware. $|\mathit{V(T,\omega)}|$, being a positive definite phasor magnitude, is then guaranteed to produce a positive resistance with a scalar offset proportional to the total SNR. For well-isolated experiments the raw $\mathit{X(T)}$ and $\mathit{Y(T)}$ signals are often observed to fluctuate around zero resistance even without accounting for any possible phase ambiguity (see Supplementary Information of Ref. \cite{salke_evidence_nodate}).

\subsection{\textbf{System Transfer Function}}
\label{sub:transfer}

\begin{figure}
\centering
  \includegraphics[width=0.4\textwidth]{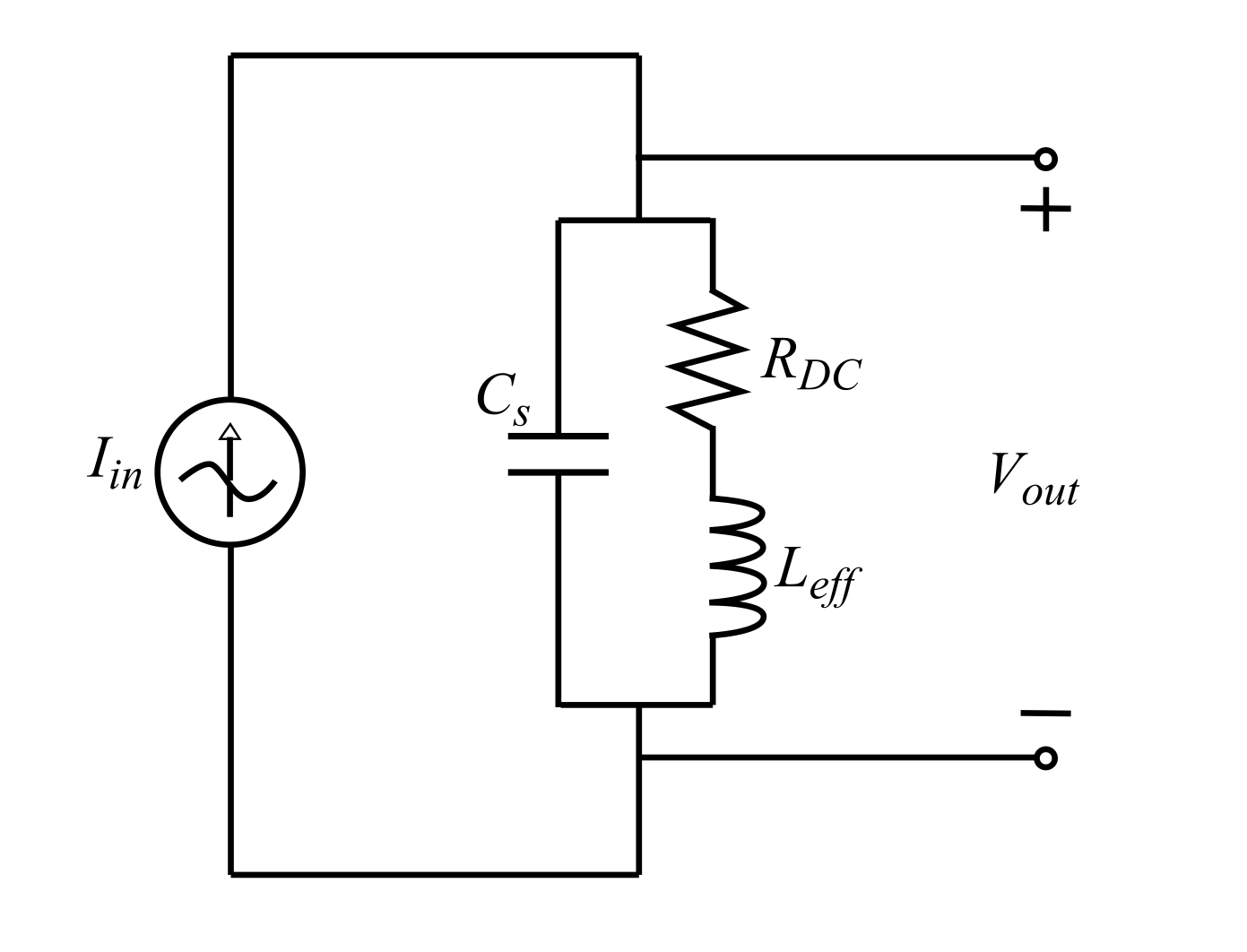}
\caption{Idealized model of a sample. $\mathit{R_{DC}}$, $\mathit{L_{eff}}$, and $\mathit{C_s}$ are in general functions of $\mathit{T}$, $\mathit{P}$, and $\mathit{\omega}$.}
\label{Fig:sample-circuit}       
\end{figure}

Modeling $\mathit{\mathcal{H}}$ of a system undergoing a complex electronic phase change, such as a granular superconductor, is achieved using passive electrical components (Fig. \ref{Fig:sample-circuit}). 

To capture the dynamics inherent to the AC measurement we choose to model the system using effective resistances, capacitances, and inductances representative of the normal state bulk properties. Previous models disregarded these dynamic effects and considered only resistive networks  \cite{raven_filamentary_1995,phillips_filamentary_1997}. In this model the effective inductance takes the form $\mathit{L_{eff}=L_{self}+L_{tr}}$ with $\mathit{L_{self}}$ being the classical self-inductance and $\mathit{L_{tr}}$ being a transient inductance present at magnetic phase changes. $\mathit{L_{self}}$ is typically highly nonlinear with closed-form solutions only existing for relatively simple boundary conditions (\textit{e.g.}, a long thin conducting wire carrying a sinusoidal current \cite{rosa_self_1908}). A finite self-capacitance ($C_{s}$) can arise from sample granularity, where adjacent grains are capacitively shunted to each other \cite{sang_interfacial_2004,wei_dielectric_2008,winkel_implementation_2020, zhai_stress-dependent_2018}. In general, $\mathit{R_{DC}}$, $\mathit{L_{eff}}$, and $\mathit{C_{s}}$ are functions of extrinsic variables, \textit{e.g.}, $\mathit{T}$, $\mathit{P}$, $\mathit{\omega}$, sample size, and sample granularity.

The admittance of the circuit (Fig. \ref{Fig:sample-circuit}) is then 
\begin{equation}
\frac{1}{\mathbf{Z}}=\frac{1}{\mathit{R_{DC}+\mathit{i}\omega L_{eff}}}+\mathit{i\omega C_{s}}
\end{equation}
which may be inverted to produce the complex transfer function (Eq. \ref{eq:tranfer-func}). \hfill
\pagebreak
\begin{strip}
\rule[-1ex]{2\columnwidth}{1pt}\rule[-1ex]{1pt}{1.5ex}
\begin{align}
\mathcal{H}(\mathit{\omega,L_{eff}} & \mathit{,C_{s},R_{DC}})= \frac{\mathit{R_{DC}^3+\omega^2R_{DC}L_{eff}^2}}{\mathit{R_{DC}^2+\omega^2[R_{DC}^4C_{s}^2+L_{eff}^2-2R_{DC}^2L_{eff}C_{s}]+\omega^4[2R_{DC}^2L_{eff}^2C_{s}^2-L_{eff}^3C_{s}]+\omega^6[L_{eff}^4C_{s}^2]}} \notag\\
& -\mathit{i\omega}\frac{\mathit{[R_{DC}^4C_{s}-R_{DC}^2L_{eff}]+\omega^2[2R_{DC}^2L_{eff}^2C_{s}]+\omega^4[L_{eff}^4C_{s}]}}{\mathit{R_{DC}^2+\omega^2[R_{DC}^4C_{s}^2+L_{eff}^2-2R_{DC}^2L_{eff}C_{s}]+\omega^4[2R_{DC}^2L_{eff}^2C_{s}^2-L_{eff}^3C_{s}]+\omega^6[L_{eff}^4C_{s}^2]}}
\label{eq:tranfer-func}
\end{align}
\hfill\rule[1ex]{1pt}{1.5ex}\rule[2.3ex]{2\columnwidth}{1pt}
\end{strip}%

$\Re\{\mathcal{H}\mathit{(t,\omega)}\}$ is the system in-phase steady-state response, which is proportional to $\mathit{X(t,\omega)}$ up to an ambiguous phase offset. $\Im\{\mathcal{H}\mathit{(t,\omega)}\}$ is the steady-state \\ quadrature response which is similarly proportional to $\mathit{Y(t,\omega)}$ up to a phase \cite{steinmetz_reactance_1894}. The phase ambiguity may be accounted for by recognizing that lock-in amplifiers inherently take differential measurements on two orthogonal quantities, so $\mathit{Y}(\mathit{t}=0)=0$ may be manually set through the use of a null detector to record only the phase change throughout the experiment. The signal magnitude and phase shift are then given by
\begin{equation}
|\mathit{V(T)|} = \sqrt{\Re\{\mathcal{H}(T)\}^2 + \Im\{\mathcal{H}(T)\}^2}
\end{equation}
and
\begin{equation}
\theta(T)=\mathrm{atan}\bigg( \frac{\Re\{\mathcal{H}(T)\}}{\Im\{\mathcal{H}(T)\}} \bigg)
\end{equation}

\begin{figure}
  \includegraphics[width=0.5\textwidth]{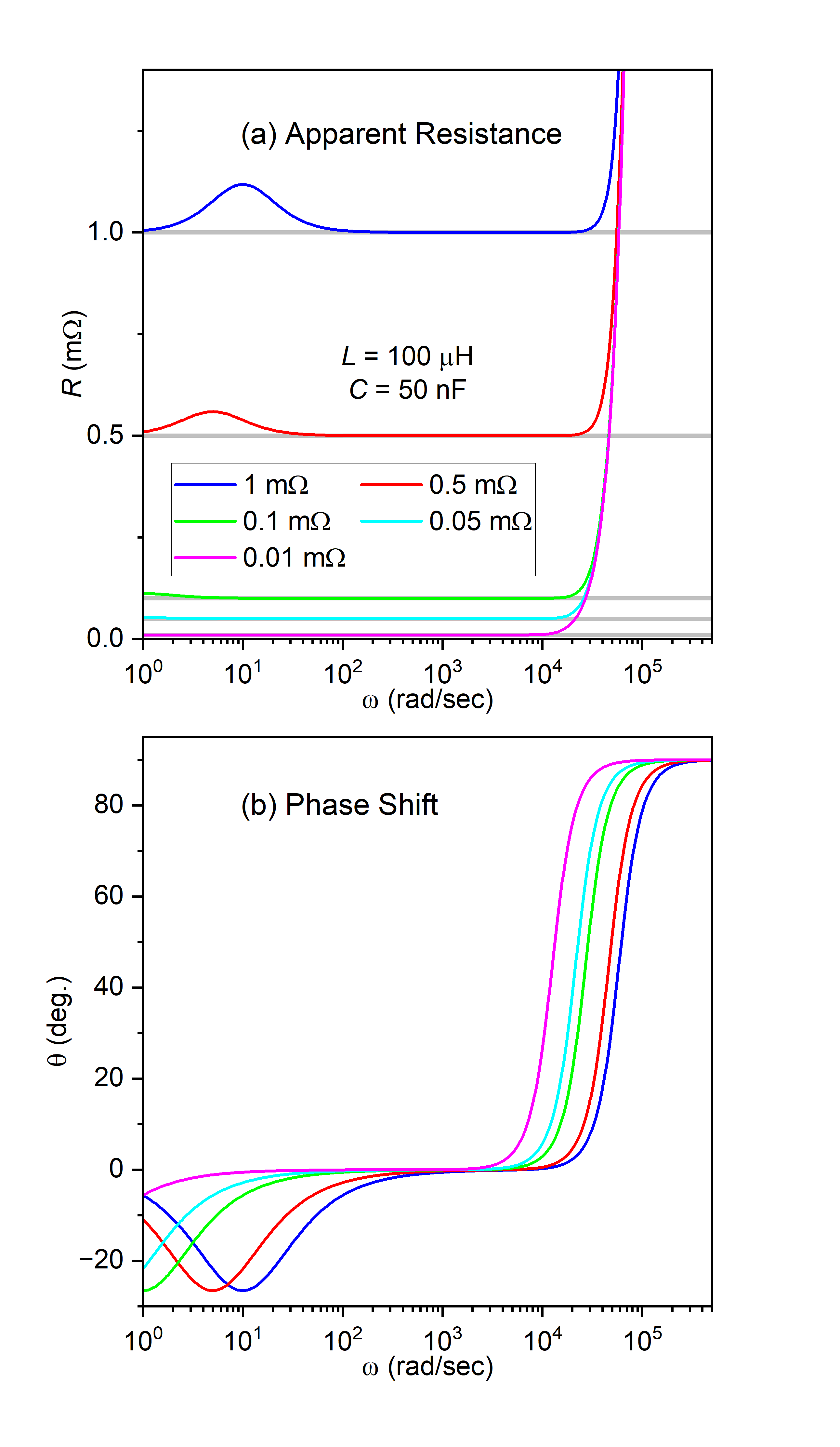}
\caption{(a) $\mathit{R_{apparent}}$ for the LRC network in Fig. \ref{Fig:sample-circuit} with constant $L$=100 $\mathrm{\mu H}$ and $C$=50 nF using various resistive loads. $\mathit{\omega}$ spans the pass-band of a typical lock-in amplifier. (b) Corresponding phase shifts for $R_{apparent}$ values.}
\label{Fig:sim-load}       
\end{figure}

We first consider the case of $\mathit{R_{apparent}}$ deviating from $\mathit{R_{DC}}$ for different resistive loads in this model with constant $\mathit{L}=1 \ \mathrm{mH}$ and $\mathit{C}=1 \ \mathrm{nF}$, values similar to those measured in DAC experiments \cite{han_electrostrictive_2021} (Fig. \ref{Fig:sim-load}a) where for $\mathit{R_{DC}}<0.5 \ \mathrm{m\Omega}$ low excitation frequencies cause $\mathit{R_{apparent}}$ to accurately reproduce the true DC resistance. Although lower frequencies are less susceptible to resonant effects they require significantly larger integration times to obtain reasonable SNR so in practice higher frequencies are typically employed. As the load and/or frequency is increased electrically reactive effects become significant and $\mathit{R_{apparent}}$ begins to diverge from $\mathit{R_{DC}}$ due to a small resonance near $\mathit{\omega_{R1}} \sim 10$, and a larger resonance above $\mathit{\omega_{R2}} \sim 10000$. Corresponding phase shifts also indicate large load-dependent effects (Fig. \ref{Fig:sim-load}b). For $\mathit{\omega_{R1} < \omega < \omega_{R2}}$, $\mathit{R_{apparent}}$ will approximate the true $\mathit{R_{DC}}$. Above $\mathit{\omega_{R2}}$ the dynamic effects may cause the measured resistance to saturate the lock-in amplifier with the signal exceeding the dynamic range of the instrument, as discussed in Ref. \cite{somayazulu_evidence_2019} (\textit{e.g.}, Fig. 4). For many experiments, $\omega_{R2}$ is above the pass-band frequency of the lock-in and is not observed. We note that if a sample undergoes a significant change in electronic properties, such as a superconducting transition, as a function of temperature, the measurement will naturally sweep through one or more of these nonlinear resonances. During the transition, $\mathit{R_{apparent}}$ is expected to diverge from $\mathit{R_{DC}}$, particularly near $\mathit{T_c}$.

\subsection{\textbf{Modeling Superconductivity}}
\label{sub:model SC}

We now model the resistance as a function of temperature and pressure with  $\mathit{R_{DC}(T,P)}$, $\mathit{L_{eff}(T,P)}$ and $\mathit{C_{s}}$ values assumed based on previously reported measurements of superconducting transitions under pressure \cite{salke_evidence_nodate} and ambient pressure measurements of Sn-whiskers \cite{miller_fluctuation_1973}, $\mathrm{2H}$-$\mathrm{NbSe_2}$ \cite{perconte_low-frequency_2020}, and $\mathrm{Bi_2Sr_2Ca_2Cu_3O_{8+\delta}}$ (Bi-2223) (see Supplemental Information and Ref. \cite{mark_notitle_nodate}). A $\mathit{R_{DC}}$-$\mathit{T}$ relation of the form 
\begin{equation}
\mathit{R(T)}=\frac{\mathit{R_{DC}(T)}}{\exp\big(\frac{\mathit{T-\mathit{T_c}}}{\Gamma}\big)-1}
\end{equation}
is assumed, where $\Gamma$ is a constant broadening factor and $\mathit{R_{DC}(T)}$ is the temperature dependence of the resistance above $\mathit{T_c}$. 

Inductive effects are magnetic and well described by Lenz's law in which any change in magnetic flux ($\Phi_\mathit{B}$) generates a proportional back electromotive force ($\mathcal{E}_{\mathit{Back}}$) determined by the sample's effective inductance $(\mathit{L_{eff}}$) \cite{graf_modern_1999}. For a sinusoidal signal, Lenz's law implies 
\begin{equation}
\mathcal{E}_{\mathit{Back}} = - \frac{d\Phi_{\mathit{B}}}{d\mathit{t}} = -|\mathit{I}|\mathit{\omega L_{eff}}
\end{equation}
with $\mathcal{E}_{Back}$ generating a voltage $90^o$ out of phase from the driving current. $\mathit{L_{tr}}$ then arises from the sudden repulsion of $\Phi_\mathit{B}$ within a superconductor at $\mathit{T_c}$ due to the Meissner effect. $\mathit{L_{tr}}$ is suppressed below $\mathit{T_c}$ due to the superconductor's inability to sustain a voltage drop. This $\mathit{L_{tr}}$ is unrelated to the kinetic inductance that has been observed in high-frequency supercurrents \cite{meservey_measurements_2003,raychaudhuri_phase_2021} which are typically several orders of magnitude smaller at the frequencies used in these experiments and only exist well below $\mathit{T_c}$ \cite{vodolazov_nonlinear_2023,liu_dynamical_2011,mondal_enhancement_2013}.   Similar transient low-frequency impedances observed in 2H-$\mathrm{NbSe_2}$ \cite{perconte_low-frequency_2020} were attributed to a coupling of the superconducting vortex state to the heat capacity. Whereas coupling to the heat capacity may cause a phase shift near $\mathit{T_c}$, it does not explain the observed frequency-dependent deviation of $\mathit{X(T)}$ from $\mathit{R_{DC}(T)}$ or the high-frequency behavior where $\mathit{R_{apparent}}$-$\mathit{T}$ can approach a step function in some measurements \cite{somayazulu_evidence_2019}. We cannot rule out variations (\textit{e.g.}, peaks) in $R_{DC}(T)$ near and above $T_c$ arising from sample or phase inhomogeneity (\textit{e.g.}, Ref. \cite{zhang_bosonic_2016}), but effects of AC measurement techniques can introduce features in this regime as well.

\begin{figure}
  \includegraphics[width=0.5\textwidth]{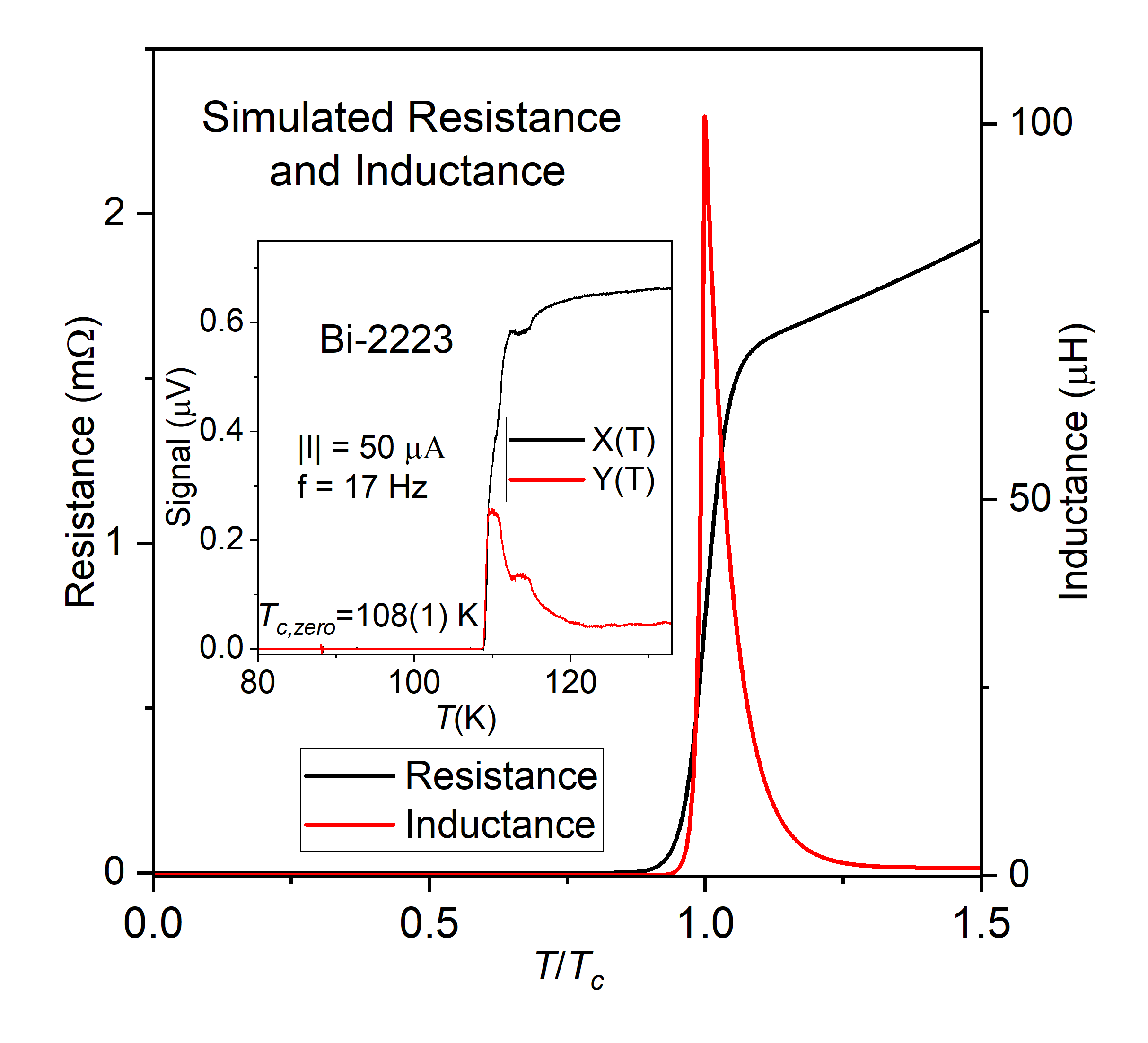}
\caption{Model $\mathit{R_{DC}}$-$\mathit{T}$ (black) and $\mathit{L_{eff}}$-$\mathit{T}$ (red) for a superconducting sample. Temperature dependence on DC resistance is assumed to be of the form $\mathit{R(T)}=\mathit{A}+\mathit{BT}+\mathit{CT^3}$ similar to the cubic relation employed in Ref. \cite{dusen_platinum-resistance_1925}. The temperature scale is normalized such that $\mathit{T_{c,mid}}$=1. The $\mathit{L_{eff}}$ lineshape is estimated based on observed $\mathit{X(T)}$ and $\mathit{Y(T)}$ signals measured in Lu-N-H \cite{salke_evidence_nodate} and cuprate samples measured here. Inset: $\mathit{X(T)}$ and $\mathit{Y(T)}$ signals from optimally doped Bi-2223 ($\mathit{T_c}$=108 K) using the setup presented in Fig. \ref{Fig:Probes}. Bi-2223 data measured over a larger temperature range are presented in the Supplementary Information.}
\label{Fig:simRL}       
\end{figure}

In our model, the total effective inductance is  described by a growing and decaying exponential surrounding $\mathit{T_c}$ modulated with a step function to account for the finite $\mathit{L_{self}}$ above $\mathit{T_c}$. Model $\mathit{R(T)}$ and $\mathit{L_{eff}(T)}$ relations that exhibit this effect are presented in Fig. \ref{Fig:simRL}. In inhomogeneous materials, including powders and polycrystals, a universal dielectric response causes an effective capacitance ($\mathit{C_{s}}$) between grain boundaries \cite{jonscher_universal_1977,almond_dielectric_1999,bouamrane_emergent_2003,almond_anomalous_2004,murphy_evidence_2006}. Eq. \ref{eq:tranfer-func} implies that lock-in based techniques will produce a $Y(T,\omega)$ that scales with $C_s$. Similar capacitive couplings between grains have been observed in ambient pressure Josephson junctions \cite{cheng_anomalous_2016,mukhopadhyay_superconductivity_2023} and in chains of capacitively coupled granular superconductors \cite{ilin_superconducting_2020}. In addition to the sample's inherent $\mathit{C_s}$, metal film contacts prepared similarly to those used in high-pressure experiments have been reported to exhibit very large contact capacitances that scale with pressure \cite{wang_stress-dependent_2021}, with some reported to exceed 1 F \cite{dervos_effect_1998}. In DAC experiments the collective capacitance of samples can be large ($C > 1 \ \mathrm{\mu}$F) and significantly impact low-frequency measurements when voltage signals are low \cite{he_situ_2007, he_alternating_2011}.  

%

%

\begin{figure}
  \includegraphics[width=0.5\textwidth]{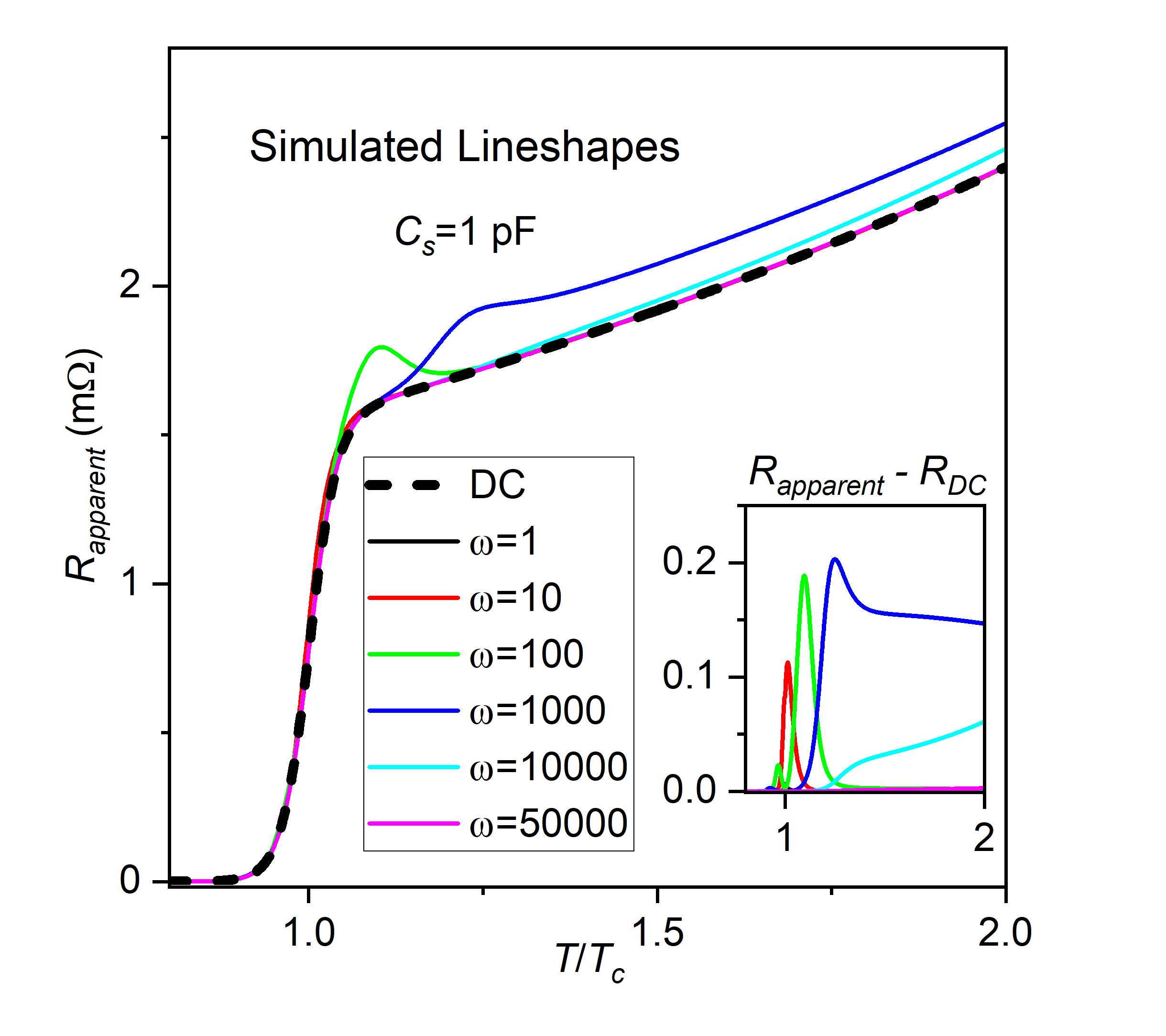}
\caption{Model $\mathit{R_{apparent}}$-$\mathit{T}$ relations for a sample with a constant $\mathit{C_{s}} = 1 \mathrm{\ pF}$ and the $\mathit{R_{DC}(T)}$ and $\mathit{L_{eff}(T)}$ relations from Fig. \ref{Fig:simRL}. Inset: Difference in $\mathit{R_{apparent}}$ and $\mathit{R_{DC}}$ for various frequencies.}
\label{Fig:transition-sim}       
\end{figure}

We now consider the frequency dependence of these effects  using the range of frequencies employed in these measurements (17 Hz to 10 kHz \cite{salke_evidence_nodate,somayazulu_evidence_2019}). Significant effects have been reported in both low frequency \cite{perconte_low-frequency_2020, salke_evidence_nodate} and high frequency \cite{somayazulu_evidence_2019} experiments. These features can provide additional information about the material in both the superconducting and normal states. As $\mathit{C_{s}}$ is relatively insensitive to temperature or frequency, it is assumed to be constant for each pressure. Model signal curves for a superconducting transition using various frequencies are presented in Fig. \ref{Fig:transition-sim}. The calculated $\mathit{R_{apparent}(T)}$ in the simulated curve does not  approximate $\mathit{R_{DC}}$ for all frequencies, especially for small samples where voltages from dynamic effects approach the resistive voltage drops. As $\mathit{\omega} \rightarrow \mathit{\omega_{R1}}$ a peak emerges in $\mathit{R_{apparent}}$ that distorts the transition (Fig. \ref{Fig:transition-sim}). While the signal is influenced by $\mathit{\omega_{R1}}$, $\mathit{R_{apparent}}$ overestimates $\mathit{R_{DC}}$ above $\mathit{T_c}$ and is seen to sharpen the transition. As $\mathit{\omega}$ is further increased, the influence of $\mathit{\omega_{R2}}$ causes the voltage signal to grow exponentially, often saturating the measurement equipment resulting in the $\mathit{R}$-$\mathit{T}$ relation approaching a step function, although this effect is not always observed due to $\mathit{\omega_{R2}}$ often being outside the range of the lock-in amplifier. Similar peaks have been reported in the resistivity of thin disordered superconductors measured using an AC bridge (\textit{e.g.}, Sn/In \cite{ems_resistance_1971} and TiN \cite{postolova_reentrant_2017}).

\begin{figure}
  \includegraphics[width=0.5\textwidth]{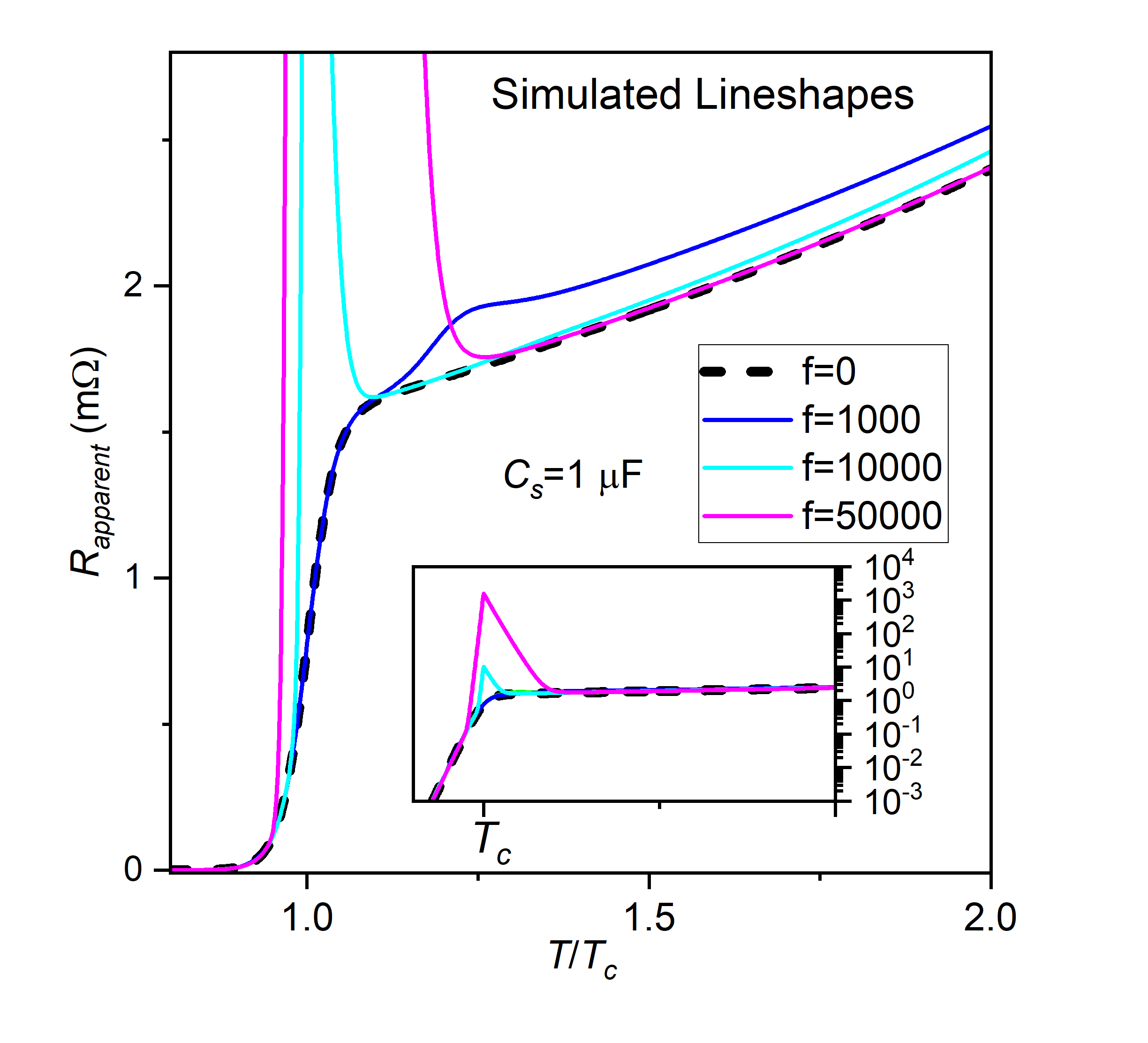}
\caption{Model $\mathit{R}$-$\mathit{T}$ relations for various high frequencies and a large $\mathit{C_{s}}$ = 1$\mu$  F. Inset: lineshape presented on a logarithmic scale demonstrating the full extent of the resonance near $\mathit{T_c}$ (see text)}
\label{Fig:large-cap}       
\end{figure}

Additional features in $R$-$T$ curves are present if the reactive contribution from $C_{s}$ is significant (\textit{e.g}. Fig. \ref{Fig:collage}b \cite{somayazulu_evidence_2019}). $C_{s}(P)$ is a function of pressure, with typical values ranging from 1 pF to 1 $\mathrm{\mu}$F \cite{wei_dielectric_2008}. This variability can cause lineshapes to be distorted under pressure (Fig. \ref{Fig:large-cap}). For larger values of $C_{s}$ and higher $\mathit{\omega}$, the system may sweep through $\mathit{\omega_{R2}}$ during the superconducting transition, producing large $X(T,\omega)$ and $Y(T,\omega)$ signals that saturate the lock-in. While significant enough to impact the measurement, the magnitude of the large resonant effects at frequencies near $\mathit{\omega_{R2}}$ is non-physical due to significant non-linear effects not accounted for in this model.

\section{Discussion}
\label{discussion}

We now discuss specific features in reported $\mathit{R}$-$\mathit{T}$ measurements of hydride high $\mathit{T_c}$ superconductors that have given rise to criticisms and concerns (\textit{e.g.}, Refs. \cite{hirsch_nonstandard_2021,hamlin_vector_nodate,dogan_anomalous_2021}) that can be understood in terms of the above considerations. First, we point out that many papers report utilizing a `van der Pauw geometry', but do not use the VDP `technique' itself (\textit{i.e.}, to measure DC sheet resistivity). Without performing a true VDP measurement resistivity cannot be determined using the VDP formula, and electrical transport properties must be determined by other means \cite{van_der_pauw_method_1958}; specifically, AC techniques imply $\mathit{R_{DC}}$ cannot be calculated using extensions of Ohm's law. Instead the full AC transfer function such as Eq. (\ref{eq:tranfer-func}) should be derived and analyzed to extract physical properties. We also point out that there is no evidence for significant broadening of the superconducting transitions shown in Fig. \ref{Fig:collage} due to pressure gradients in the sample (\textit{e.g.}, Ref. \cite{deemyad_dependence_2003}). This observation is consistent with the expected relaxation of shear stresses during synthesis and further suggests that shear stresses in the hydride materials are small \cite{somayazulu_evidence_2019}.

We now discuss examples of various features observed in specific high $\mathit{T_c}$ hydride materials. As discussed above, in the original study that led to the discovery of superconductivity in $\mathrm{LaH_{10}}$ \cite{somayazulu_evidence_2019} the sharpness of the transition and slope of the measured $R$-$T$ response depend on the properties of the measurement. Regular sample geometries produced $\mathit{R}$-$\mathit{T}$ curves reminiscent of the expected $\mathit{R_{DC}}$ drop (Fig. \ref{Fig:collage}a)  \cite{somayazulu_evidence_2019}. The same study found that irregularly shaped inhomogeneous samples, determined using spatially resolved \textit{in-situ} x-ray diffraction, exhibited different behavior. For these samples the $\mathit{R_{apparent}}$-$\mathit{T}$ curves approached a step function (Fig. \ref{Fig:collage}b). This behavior can readily be explained as arising from filtering effects due to a geometrically enhanced $\mathit{C_{s}}$ (Fig. \ref{Fig:collage}b). The result of a subsequent study \cite{drozdov_superconductivity_2019} show peaks within the resistance drop near $\mathit{T_c}$ (Fig. \ref{Fig:collage}c), features that were attributed to competing superconducting phases and metal-insulator interface effects \cite{zhang_bosonic_2016}. Similar sharp step-like behavior in $\mathrm{CeH_{9,10}}$ was observed in $R_{apparent}$-$T$ upon increasing $\omega$ \cite{semenok_evidence_nodate} (Fig. \ref{Fig:collage}d). We point out that inhomogeneity could increase sample $C_{s}$ and enhances AC resonance effects at these temperatures.


Other interesting features are apparent in the $\mathit{R}$-$\mathit{T}$ curves of compressed $\mathrm{H_3S}$ reportedly measured using an AC bridge circuit ($\mathit{f}$=13 Hz) (Fig. \ref{Fig:collage}e) \cite{mozaffari_superconducting_2019}. Peaks characteristic of the resonance at $\mathit{\omega_{R1}}$ are apparent in the resistance drop of the 155 GPa runs but are absent in the 160 GPa data. It was reported that these samples were prepared using different methods; notably the 160 GPa sample being annealed to increase homogeneity whereas that studied at 155 GPa was not. It is expected that more inhomogeneous samples would have a larger $C_{s}$ and therefore exhibit larger dynamic signals near $\mathit{T_c}$.

Reported $\mathit{R}$-$\mathit{T}$ curves for compressed C-S-H measured using a lock-in-based setup exhibit many of the above features associated with AC measurements (Fig. \ref{Fig:collage}f) \cite{hirsch_enormous_2023}; see also Ref. \cite{pasan_observation_nodate}. The curves show sharp transitions at $\mathit{T_c}$ representative of the large $\mathit{Y(T)}$ signal present at $\mathit{T_c}$. Additionally, several experimental runs exhibit broad peaks or `ripples' above $\mathit{T_c}$, particularly the data for run 1 (blue boxes) at 174 GPa and 220 GPa. As these effects are highly dependent on sample geometry it is expected that repeated runs on the same samples would produce similar features.

Superconductivity in the Lu-N-H system near ambient conditions remains a subject of continued investigations. There is evidence of superconductivity in the vicinity of room temperature and pressure in selected Lu-N-H samples \cite{dias_observation_2023,salke_evidence_nodate}. The $\mathit{R}$-$\mathit{T}$ relations measured at various pressures also exhibit narrow superconducting transitions (Fig. \ref{Fig:collage}g). Measurements at 1.6 and 2.0 GPa in Ref. \cite{dias_observation_2023} were performed on significantly larger samples than the 1.0 GPa measurements. As a result, the normal state resistance is over 100x larger for the higher pressure measurements. The smaller sample gave rise to a sharper transition, an effect that we ascribe to parasitic impedances.

\section{Conclusions}
\label{conclusions}

We have shown that classical electrodynamics, circuit theory, and plausible values for the bulk electrical properties of the materials can explain features in reported $R$-$T$ measurements of high $T_c$ hydride superconductors under pressure. In particular, we demonstrate that lineshapes in reported AC-based transport measurements that have been pointed out to be anomalous may be explained by considering all signal-processing effects. Many of the features observed in these measurements are expected for a superconducting transition. Sample dependent geometric effects are expected to alter the measured signals,\textit{e.g.}, a larger $C_s$ in an inhomogeneous material is expected to drastically enhance the measured $Y(T,\omega)$ signal at all frequencies. While instrumental and dynamic electrical effects may result in $\mathit{R_{apparent}}$ diverging from $\mathit{R_{DC}}$, particularly near and above $\mathit{T_c}$, the physicality of the superconducting transition is unambiguous. While these results are fully consistent with superconductivity, additional measurements using different techniques are, of course, needed to confirm the materials are superconductors. To accurately interpret results and avoid confusion all aspects of the measurements, including the experimental setup and relevant experimental parameters, should be reported (\textit{e.g.}, $\mathit{X}$, $\mathit{Y}$, $|\mathit{V}|$ and $\theta$). Analysis of these data can provide additional information about the state of potential superconducting samples.

\begin{acknowledgements}
We thank A. Denchfield, C. Mark, P. Melnikov, J.C. Campuzano, S. Deemyad, M. Somayazulu, D. Semenok, and G.W. Collins for helpful discussions. This work was supported by the U.S. National Science Foundation (DMR-2104881), DOE-NNSA (DE-NA0003975 Chicago/DOE Alliance Center), and DOE-SC (DE-SC0020340).
\end{acknowledgements}

\begin{Data availability}
\small{\noindent \textbf{Data Availability} All data are available upon request.} 
\end{Data availability}

\printbibliography

\pagebreak
\clearpage
\begin{refsection}
\renewcommand{\figurename}{Fig. S\!\!}
\setcounter{figure}{0}
\renewcommand{\tablename}{Table S\!\!} 
\setcounter{table}{0}
\renewcommand{\theequation}{S\arabic{equation}}
\setcounter{equation}{0}

\renewcommand{\refname}{Supplemental References}

\newrefcontext[labelprefix=S]

\begin{strip}
\begin{center}
{\large Supplemental Information}
\end{center}
\end{strip}

\section*{Lock-in Amplification}

In general, phase sensitive detection exploits the fact that many periodic waveforms are orthogonal up to a phase (see Refs. \cite{michels_pentode_1941,stutt_low-frequency_1949}). In this work, we use sinusoidal waveforms which have the orthogonality relation,

\begin{align}
\int_{0}^{2 \pi} \mathrm{sin}(\mathit{\omega_n \, s} + \phi_\mathit{n}) & \: \mathrm{sin}(\mathit{\omega_m \, s} + \phi_\mathit{m}) \: d\mathit{s} \notag\\ = & \begin{cases} \mathrm{cos}(\phi_\mathit{n}-\phi_\mathit{m}) : \mathit{\omega_n=\omega_m} \\ 0: \mathit{\omega_n \neq \omega_m} \end{cases}
\label{eq:orth}
\end{align}

\noindent where $\mathit{\omega_{n,m}}$ are frequencies and $\phi_\mathit{n,m}$ are phase shifts. This relation allows for the isolation and detection of small periodic signals, $\mathit{A(t,\omega)}$. Even if the measured signal is noisy due to environmental or instrumental noise, by modulating it with a reference signal of known amplitude and $\mathit{\omega}$ then integrating over at least one period, $\mathit{A(t,\omega)}$ can be recovered. In practice, the SNR can be increased further by integrating over many periods. The orthogonality of the sinusoidal carrier wave `averages out' any component of $\mathit{A(t,\omega)}$ that is not oscillating at $\mathit{\omega}$. The amplifier then outputs the root mean square (RMS) of the signal that is in phase with the reference sinusoid as

\begin{equation}
\mathit{X(t,\omega)}=\frac{2}{\mathit{\tau}}\int_{\mathit{t}-\tau}^{\mathit{t}} \mathit{A}(\mathit{s},\phi_{\mathit{A}}) \ \mathrm{sin}(\mathit{\omega s} + \phi_{\mathit{ref}})d\mathit{s},
\label{eq:in-phase}
\end{equation}

\noindent where $\tau$ is an integration time, $\phi_{\mathit{ref}}$ is the reference signal phase, and $\phi_{\mathit{A}}$ is the signal phase. The phase difference is then $\theta(\mathit{t})=\phi_{\mathit{A}}-\phi_{\mathit{ref}}$. Typically $\phi_{\mathit{ref}}$ is adjusted manually through a unity gain null detector in the lock-in input to ensure $\mathit{Y}(\mathit{t}=0,\mathit{\omega})=0$ before measurements are taken. 

As $\phi_{\mathit{ip}}$ might drift throughout the experiment, whether due to sample properties or parasitic effects, a quadrature measurement ($90^o$ out of phase from the reference signal) can also be measured and used to remove any phase ambiguity. This quadrature signal is given as

\begin{equation}
\mathit{Y(t,\omega)}=\frac{2}{\mathit{\tau}}\int_{\mathit{t}-\tau}^{\mathit{t}} \mathit{A}(\mathit{s},\phi_{\mathit{A}}) \ \mathrm{cos}(\mathit{\omega s} + \phi_{\mathit{ref}})d\mathit{s},
\label{eq:quad}
\end{equation}

\noindent For $\tau >> \frac{2\pi}{\mathit{\omega}}$ equations \ref{eq:in-phase} and \ref{eq:quad} reduce to 
\pagebreak

\begin{equation}
\mathit{X(t,\omega)} \approx |\mathit{A_{RMS}(t,\omega)}| \cos[\theta\mathit{(t,\omega)}]
\end{equation} 
and 
\begin{equation}
\mathit{Y(t,\omega)} \approx |\mathit{A_{RMS}(t,\omega)}| \ \sin[\theta\mathit{(t,\omega)}],
\end{equation}
respectively. To relate the measurements to physical values,  output signals can be converted to polar coordinates and root mean square values can be converted into amplitudes, \textit{e.g}, multiplying by $\sqrt{2}$ for sinusoidal carrier waves. Being mutually orthogonal, $\mathit{X(t,\omega)}$ and $\mathit{Y(t,\omega)}$ can be defined on a complex plane to give

\begin{equation}
\mathit{\mathbf{A}(t,\omega)}=\mathit{X(t,\omega)}+\mathit{iY(t,\omega)} =  |\mathit{A(t,\omega)}|e^{\mathit{i}\theta(\mathit{t,\omega})}.
\label{eq:polar}
\end{equation}

\noindent $|\mathit{A(t)}|$ is then the output signal magnitude and $\theta\mathit{(t)}$ is the phase offset from the lock in reference signal. In this work $|\mathit{A}|$ is used to denote the output magnitude as opposed to the more common $\mathit{R}$ to avoid any ambiguity with resistance. Throughout this discussion, it is important to keep in mind that the lock-in amplifier only measures $\mathit{X(t,\omega)}$ and $\mathit{Y(t\omega)}$ which are used to calculate 
\begin{equation}
|\mathit{A(t,\omega)}|=\sqrt{[\mathit{X(t,\omega)}]^2 + [\mathit{Y(t,\omega)}]^2}
\end{equation}
and 
\begin{equation}
\theta(\mathit{t,\omega})=\mathrm{arctan} \big(\frac{\mathit{Y(t,\omega)}}{\mathit{X(t,\omega)}}\big)
\end{equation}
with all instrumental noise and uncertainty propagated accordingly.

\section*{Fault Detection}
\begin{figure}
\center
  \includegraphics[width=0.5\textwidth]{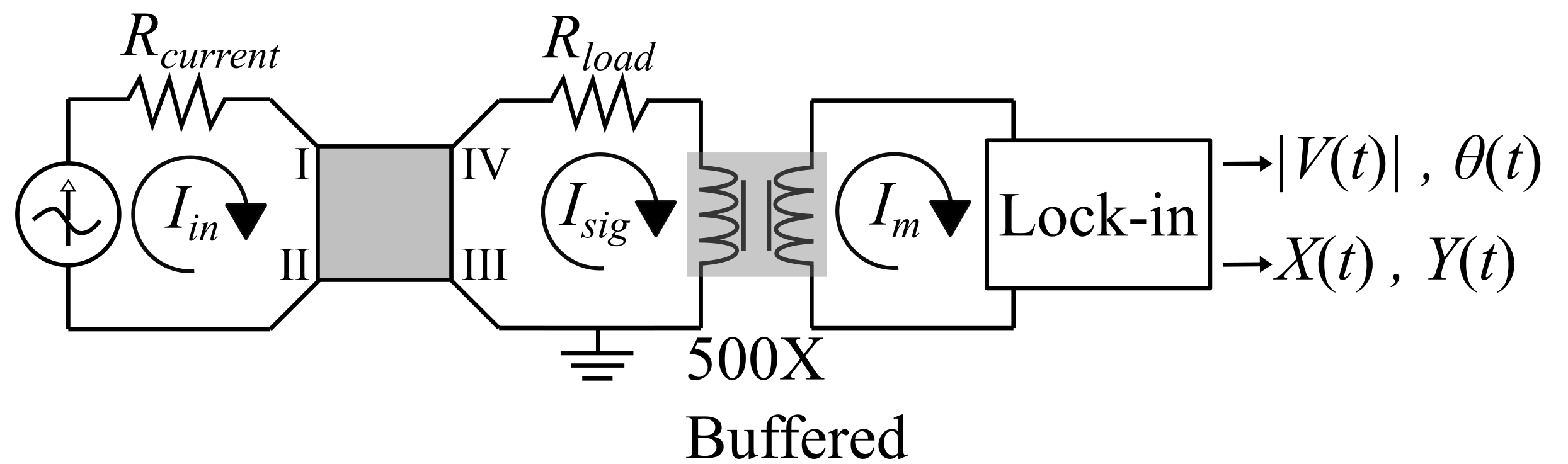}
\caption{Example of the schematic block diagram of a setup used for small-signal transport measurements with the sample grounded. Small parasitic resistances in the current and voltage lines are represented by $\mathit{R_{current}}$ and $\mathit{R_{load}}$, respectively. Grounding the sample may result in fictitious zero resistance readings, and such a circuit topology should not be employed for low-resistance measurements.}
\label{Fig:Real-circuit-ground}       
\end{figure}

\begin{figure}
\center
  \includegraphics[width=0.5\textwidth]{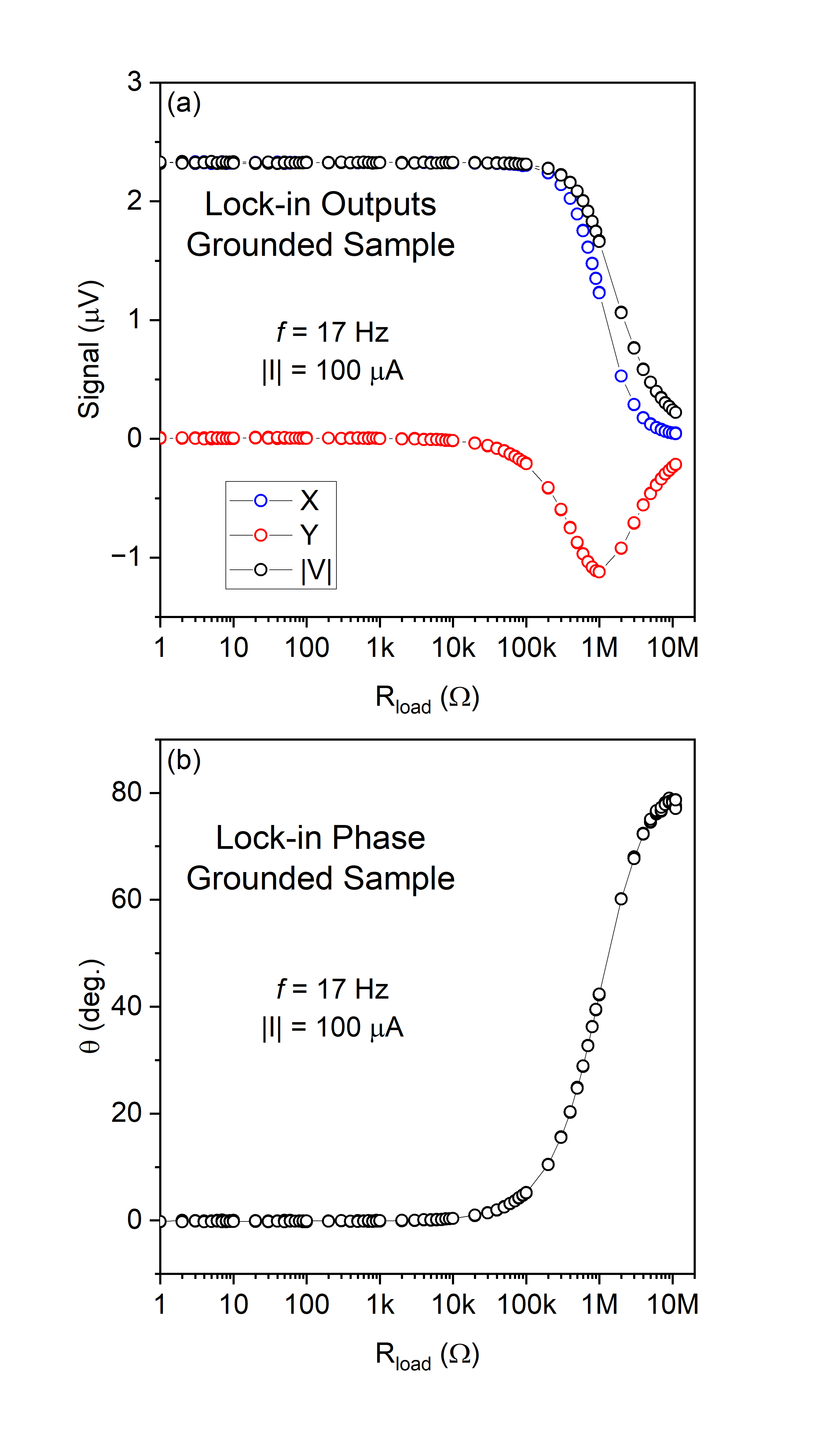}
\caption{a) Voltage outputs for a grounded circuit (Fig. S\ref{Fig:Real-circuit-ground}) measuring a sheet of copper. $X$, $Y$ and $|V|$ as functions of $R_{load}$ (an in-line variable resistor) are plotted on a log scale. b) $\theta$ response to loading effects.}
\label{Fig:Ground-XY}      
\end{figure}

\begin{figure}
\center
  \includegraphics[width=0.5\textwidth]{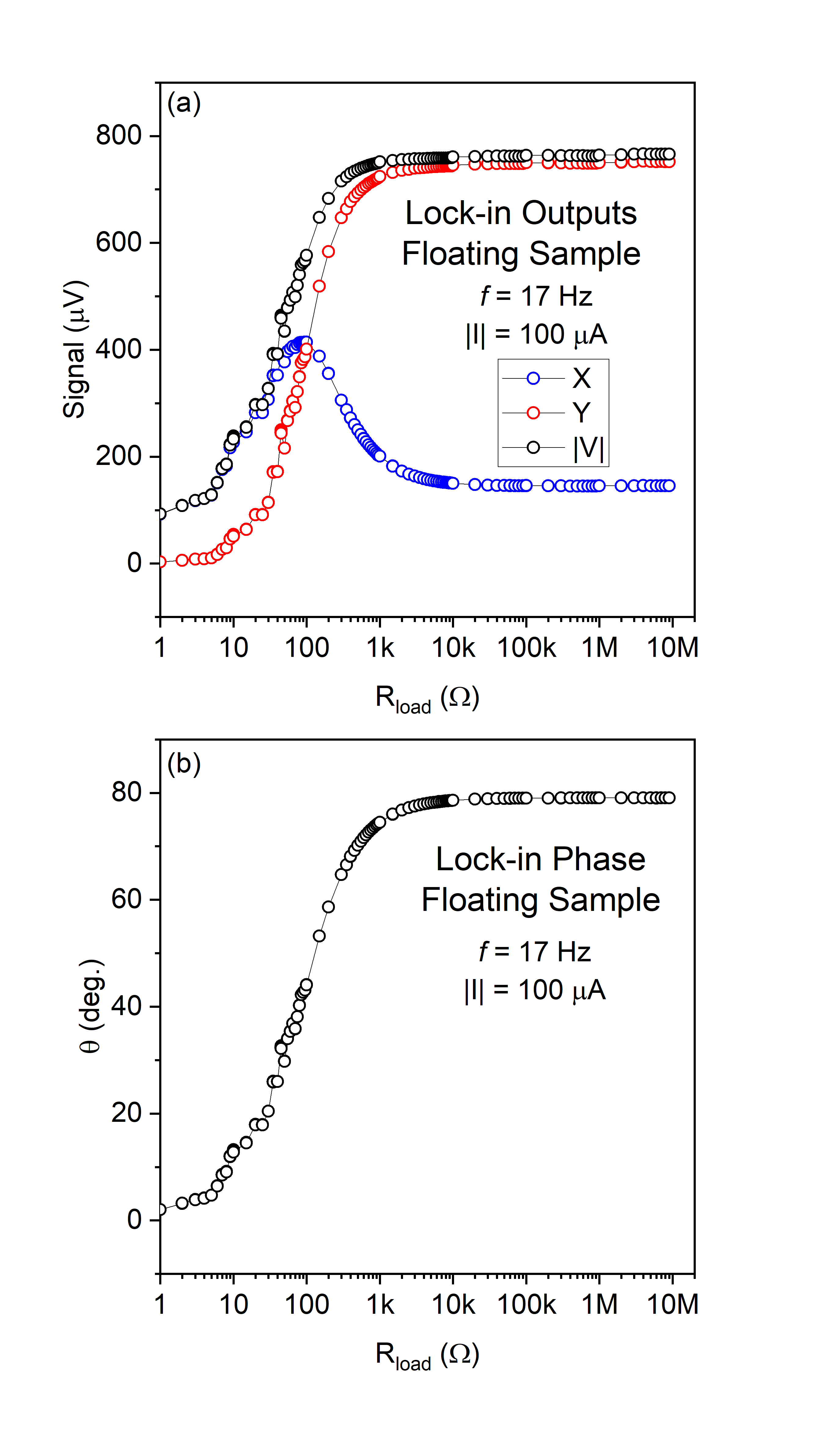}
\caption{a) Voltage outputs for a floating circuit (main text Fig. 2) measuring a sheet of copper. $X$, $Y$ and $|V|$ as functions of $R_{load}$ (an in-line variable resistor) are plotted on a log scale. b) $\theta$ response to loading effects.}
\label{Fig:float-XY}      
\end{figure}

In order to prevent pre-amplifier loading effects, the sample was electrically floating (main text Fig. 2). In a grounded sample with inhomogeneities in the electrical resistance (Fig. S\ref{Fig:Real-circuit-ground}) the current may bypass the buffered pre-amplifier entirely and short to ground, producing a zero-voltage signal for an ohmic material (Fig. S\ref{Fig:Ground-XY}). In order to ensure the signal is not shunted to ground the sample is electrically floating and only a differential voltage is measured (main text Fig. 2). Removing the path to ground prevents fictitious zero voltage measurements in the event of large in-line resistances or contact breakage. Increasing $R_{load}$ in the floating circuit produces increasing signal values and causes $\theta$ to saturate near 80$^o$ (Fig. S\ref{Fig:float-XY}).

%

%

\section*{Bi-2223 Transport Data}

\begin{figure}
\center
  \includegraphics[width=0.5\textwidth]{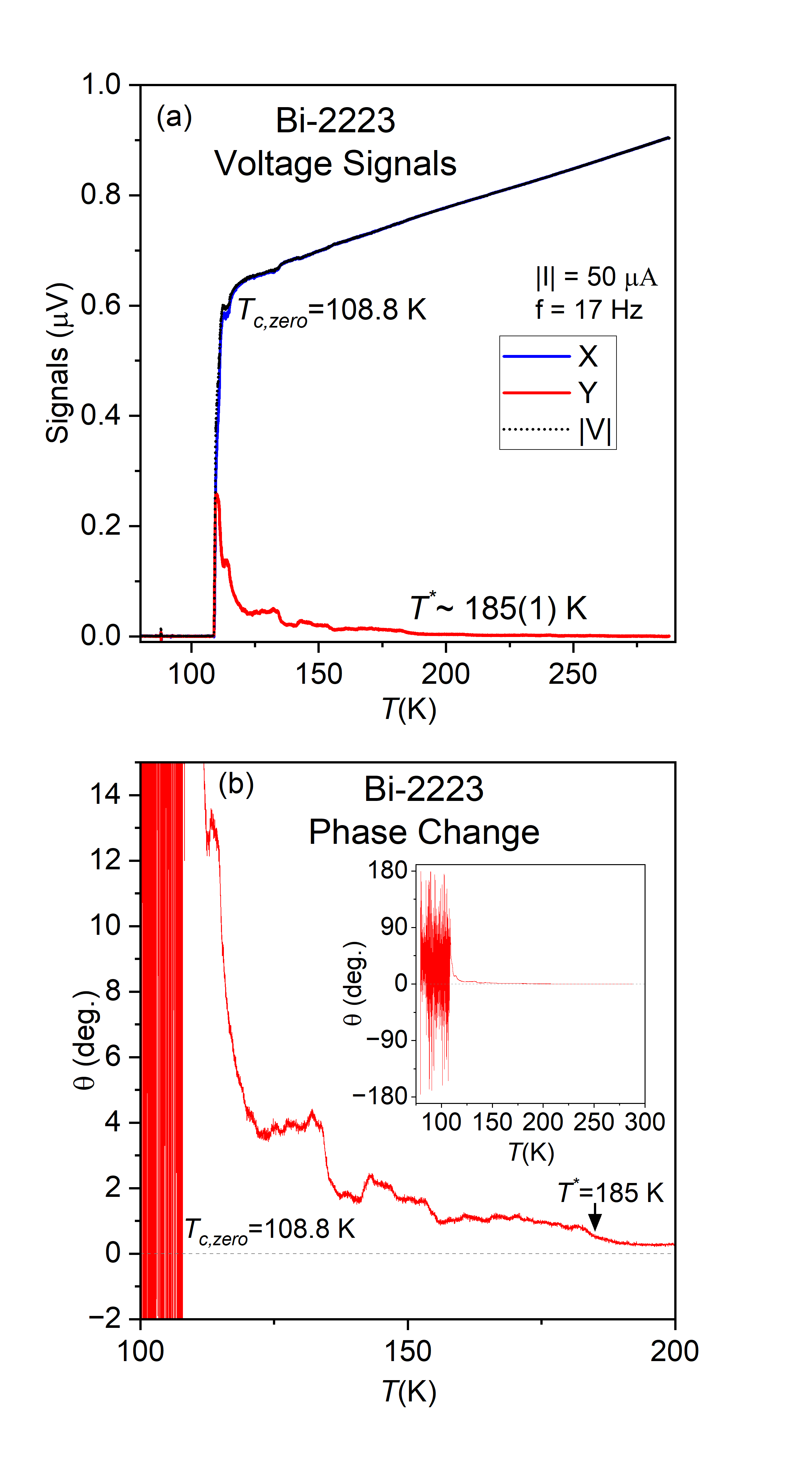}
\caption{a) Bi-2223 lock-in signals vs temperature measured using the circuit presented in Fig. 2. b) Corresponding phase angle vs temperature from 100 K - 200 K. There is a shift in the phase angle at 185 K that is attributed to the pseudogap onset ($\mathit{T^*}$). Below $\mathit{T_c}$ the signal is chaotic due to both $\mathit{X(T)}$ and $\mathit{Y(T)}$ being noise resolved. Inset: Measurement over full temperature range.}
\label{Fig:2223-signals}       
\end{figure}

Ambient pressure electrical transport data from sintered bars of Bi-2223 ($\mathit{T_c} = 108 \ K$) purchased from Quantum levitation. Measurements were taken using the measurement circuit presented in Fig. \ref{Fig:Probes}. Voltage and phase signals as functions of temperature are presented in Fig. S\ref{Fig:2223-signals}. \vfill

\pagebreak
\newpage
\printbibliography
\end{refsection}
\end{document}